\documentclass[12pt]{article}  

 \usepackage{amsmath}

\title{Uncertainty Relations and Entanglement for PQ-deformed Supersymmetric Coherent States}

\author{Oktay K Pashaev and Aygul Kocak\\
Department of Mathematics \\ Izmir Institute of Technology \\ Izmir 35430, Turkey}

\begin{document}

\maketitle              

\begin{abstract}
     We propose supersymmetric extension of deformed quantum oscillator with two parameters quantum group structure. As particular cases, specified by values of $p$ and $q$ parameters it includes symmetric and
non-symmetric $q$-oscillators, Fibonacci and Fibonacci divisors hierarchy of Golden oscillators, Tamm-Dankov oscillator etc. By $pq$-deformed supersymmetric annihilation operator, the set of 
corresponding supersymmetric coherent states is introduced. The states are characterized by the pair of $pq$-quantum states from the Fock space or equivalently, by the set of infinite number of
qubit states. Entanglement of fermions with $pq$-deformed bosons is characterized by the concurrence as the linear entropy,  taking form of the Gram determinant of inner products. As shown, for two types of the reference states, the concurrence depends on values of $p$ and $q$ parameters, and it reaches the value one for the maximally entangled states.  Entanglement of the super-coherent states and the uncertainty relations for the coordinate and momentum in these states are calculated. Non-classical nature of the entangled states is reflected in non-minimal character of the uncertainty relations.
\end{abstract}



 %
\section{Introduction}
  The $pq$-deformed extensions of quantum oscillator with two base parameters were developed for general values of $p$ and $q$ in \cite{CJ}, \cite{A}, and for specific values of the parameters 
as the Golden calculus in \cite{P1}, and the hierarchy of Golden oscillators in \cite{P4}. More general deformation of quantum mechanics and quantum statistics by two parameter quantum group recently was discussed in 
\cite{AC}. Extension of quantum oscillator to the supersymmetric case shows the novel property of entanglement between bosonic and fermionic degrees of freedom \cite{PA}, which influences 
coordinate-momentum uncertainty relations in corresponding supersymmetric coherent states.  Entanglement property for Golden bosonic oscillator and fermions in supersymmetric coherent states was studied in \cite{P24}. In the present paper we describe entanglement and uncertainty relations for supersymmetric extension of general two-parameter deformed quantum oscillator.

 \section{The $pq$-Quantum Calculus} 
	The $pq$ - number is defined as
	\begin{equation}
	[n]_{p,q} = \frac{p^n - q^n}{p-q} = [n]_{q,p}\,.
	\end{equation}
	For different values of parameters $p$ and $q$ it includes following particular cases.
	
	{\bf Example:} 
	
	Non-symmetric $q$-numbers:
	 $p=1$, 
	\begin{equation}
	[n]_{q,1} \equiv [n]_q = \frac{q^n - 1}{q-1}\,.
	\end{equation}

	{\bf Example:} 
	
	Symmetric $q$ numbers:
	$p = q^{-1}$, 
	\begin{equation}
	[n]_{q, \frac{1}{q}} = \frac{q^n - q^{-n}}{q - q^{-1}}\,.
	\end{equation}

	{\bf Example:} 
	
	Fermionic $q$ numbers:
	$p = - q^{-1}$,
	\begin{equation}
	[n]_{q, -q^{-1}} = \frac{q^n - (-1)^n q^{-n}}{q + q^{-1}}\,.
	\end{equation}

	{\bf Example:} 
	
	Fibonacci numbers:
$p = \varphi = (1+ \sqrt{5})/2$ - the Golden ratio, $q = \varphi' = - \varphi^{-1}$, 
	\begin{equation}
	[n]_{\varphi, \varphi'} = F_n = \frac{\varphi^n - \varphi'^n}{\varphi - \varphi'}\,.
	\end{equation}

	{\bf Example:} 
	
	Fibonacci divisor numbers:
	$p = \varphi^k$, $q = \varphi'^{k}$, 
	\begin{equation}
	[n]_{\varphi^k, \varphi'^k} = \frac{{\varphi^k}^n - {\varphi'^k}^n}{\varphi^k - \varphi'^k} = \frac{F_{n k}}{F_k}\,.
	\end{equation}

	{\bf Example:} 
	
	Tamm-Dankov numbers:
	$p = q$, 
	\begin{equation}
	[n]_{q, q} = n q^{n-1}\,.
	\end{equation}
	 
	\subsection{The algebraic properties}
The numbers satsfy following algebraic relations
\begin{eqnarray}
	[n+1]_{p,q} &=& q [n]_{p,q} + p^n = p [n]_{p,q} + q^n\,, \\
	{[n+m]_{p,q}} &=& p^n {[m]_{p,q}} + q^m {[n]_{p,q}}\,, \\
	{[n-m]_{p,q}} &=& p^n {[-m]_{p,q}} + q^{-m} {[n]_{p,q}}\,, \\
	{[-n]_{p,q}} &=&  -{ \frac{[n]_{p,q}}{(p q)^n}}\,, \\
	{[n-m]_{p,q}} &=& q^{-m} {( [n]_{p,q} - p^{n-m} [m]_{p,q} )}\,, \\
	{[n m]_{p,q}} &=& {[n]_{p,q}} {[m]_{p^n,q^n}}  = {[m]_{p,q}} {[n]_{p^m,q^m} }\,,\\
	\left[\frac{n}{m}\right]_{p,q} &=& \frac{[n]_{p,q}}{[m]_{p^{\frac{n}{m}}, q^{\frac{n}{m}}}} = \frac{[n]_{p^{\frac{1}{m}},q^{\frac{1}{m}}}}{[m]_{p^{\frac{1}{m}}, q^{\frac{1}{m}}}}\,.
	\end{eqnarray}
	The base inversion and factorial formulas are
	\begin{eqnarray}
	[n]_{p,q} &=& [n]_{\frac{1}{p}, \frac{1}{q}} (p q)^{n-1}\,, \\ {[n]_{p,q} !} &=& {[n]_{\frac{1}{p}, \frac{1}{q}} !} (p q)^{\frac{n(n-1)}{2}}\,.
	\end{eqnarray}
	
	\subsection{The $pq$ numbers as generalized Fibonacci numbers}
	The deformed numbers satisfy three-term recursion formula
	\begin{equation}
	[n+1]_{p,q} = (p+q) [n]_{p,q} - p q [n-1]_{pq}\,,
	\end{equation}
	generalizing the one for Fibonacci numbers.
	
	\subsection{The $pq$-calculus}
The $pq$-derivative is defined as
\begin{equation}
D_{p,q} f(z) = \frac{f(pz) - f(qz)}{(p-q) z} = D_{q,p} f(z)\,,
\end{equation}
with following action  on monomial 
\begin{equation}
D_{p,q} z^n = [n]_{p,q} z^{n-1}
\end{equation}
and on $pq$-binomial, 
\begin{equation}
D_{p,q} (z - a)_{p,q}^n = [n]_{p,q} (z-a)^{n-1}_{p,q},
\end{equation}
where
\begin{equation}
(z-a)^n_{p,q} = (z- p^{n-1} a) (z - p^{n-2} q a) ...(z - p q^{n-2}) (z- q^{n-1} a)
\end{equation}
for $n \ge 1$, and it is equal one for $n=0$. 
The deformed binomial satisfies relations
\begin{eqnarray}
(z-a)_{p,q}^{n+m} = (z-p^m a)^n_{p,q} (z - q^n a)^m_{p,q} = (z-q^m a)^n_{p,q} (z - p^n a)^m_{p,q},
\end{eqnarray}
\begin{eqnarray}
(z - q^n a)^{-n}_{p,q} = \frac{1}{(z - p^{-n} a)^n_{p,q}} \,,
\end{eqnarray}
\begin{eqnarray}
(z + a)^n_{p,q} = \sum^n_{k=0} \left[ \begin{array}{c} n \\ k        \end{array}\right]_{p,q} (p q)^{\frac{k(k-1)}{2}} z^{n-k} a^k\,.
\end{eqnarray}
	
	\subsubsection{The $pq$-exponential functions}
Two types of the $pq$-exponential functions 
\begin{equation}
e^z_{p,q} = \sum^{\infty}_{n=0} \frac{z^n}{[n]_{p,q} !},\hskip1cm E^z_{p,q} = \sum^{\infty}_{n=0} \frac{z^n}{[n]_{p,q} !} (pq)^{\frac{n(n-1)}{2}}\,,
\end{equation}
are connected by formula
\begin{equation}
e^z_{p,q} = E^z_{\frac{1}{p},\frac{1}{q}}\,,
\end{equation}
and satisfy equations
\begin{equation}
D_{p,q} e^z_{p,q} = e^z_{p,q}, \hskip1cm D_{p,q} E^z_{p,q} = E^{p q z}_{p,q}\,.
\end{equation}
	
	\subsubsection{\bf Proposition.}
For $p$ and $q$ real, so that $|q| > 1$ and $0 < |p| < |q|$, function $e^z_{p,q}$ is entire analytic function of $z$,
satisfying relation
\begin{equation}
e^{p z}_{p,q} - e^{q z}_{p,q} = (p-q) z e^{z}_{p,q} \,.\label{pqrelation}
\end{equation}

\subsubsection{Proposition.}
Function 
\begin{equation}
f(\lambda, z) \equiv \frac{e^{\lambda z}_{p,q}}{e^{ z}_{p,q}} - \lambda z \label{f}
\end{equation}
satisfies relations
$f(p, z) = f(q, z)$, $f(1, z) = 1-z$.

	\section{The pq-deformed Quantum Oscillator}

  Here, we briefly recall basic properties of  two parametric, the $pq$-deformed quantum oscillator, introduced in  \cite{CJ}, \cite{A}. It is determined by the pair of creation and annihilation operators, $a^\dagger_{p,q}$, $a_{p,q}$, and the number operator $N$, satisfying the quantum algebra
\begin{eqnarray}
a_{p,q} a^\dagger_{p,q} - p a^\dagger_{p,q} a_{p,q} = q^{N}, \\
a_{p,q} a^\dagger_{p,q} - q a^\dagger_{p,q} a_{p,q} = p^{N},  
\end{eqnarray}
$[N,a^\dagger_{pq}] = a^\dagger_{pq}$ and    $[N, a_{pq}] = -a_{pq}$. 
The pq-number operator is determined by formula
\begin{equation}
[N]_{p,q} = \frac{p^N - q^N}{p-q}
\end{equation}
and it can be represented 
in factorized form as
\begin{eqnarray}
[N]_{p,q} = a^\dagger_{p,q} a_{p,q}\,, \hskip1cm [N+I]_{p,q} = a_{p,q} a^\dagger_{p,q}\,.
\end{eqnarray}
The operator satisfies following relations
\begin{eqnarray}
[N+I]_{p,q} &=& (p+q) [N]_{p,q} - pq [N-I]_{p,q}, \\
(p)^N &=& p [N]_{p,q} - p q [N-I]_{p,q}, \\
(q)^N &=& q [N]_{p,q} - p q [N-I]_{p,q}, \\
{[N+I]_{p,q}} &=& p {[N]_{p,q}} + q^{ N}  = q {[N]_{p,q}} + p^{ N},
\end{eqnarray}
gives expression for commutator
\begin{eqnarray}
    a_{p,q} a^\dagger_{p,q} - a^\dagger_{p,q} a_{p,q} = [a_{p,q}, a^\dagger_{p,q}] = [N+I]_{p,q} - [N]_{p,q},
\end{eqnarray}
and
\begin{eqnarray}
[[N]_{p,q}, {a_{p,q}^\dagger}] &=& ([N]_{p,q} - [N-I]_{p,q}) {a_{p,q}^\dagger} = {a_{p,q}^\dagger} ([N+I]_{p,q} - [N]_{p,q}) , \\
a_{p,q}^\dagger f([N+I]_{p,q}) &=& f([N]_{p,q}) a_{p,q}^\dagger, \hskip0.5cm
a_{p,q} f([N]_{p,q}) = f([N+I]_{p,q}) a_{p,q}.
\end{eqnarray}
	
	 The Fock space basis states  $\{ |n\rangle_{p,q}   \}$, defined as
\begin{equation}
|n\rangle_{p,q} = \frac{(a^\dagger_{p,q})^n}{\sqrt{[n]_{p,q}!}} |0\rangle_{p,q},\hskip1cm a_{p,q} |0\rangle_{p,q} =0,\label{deformedFock}
\end{equation}
are the orthonormal $_{p,q}\langle n| m\rangle_{p,q} = \delta_{n m}$ eigenstates of 
$[N]_{p,q}$,
\begin{eqnarray}
[N]_{p,q} |n\rangle_{p,q} = [n]_{pq} |n\rangle_{p,q}\,, 
\end{eqnarray}
 and satisfy relations
\begin{equation}
   a^\dagger_{p,q} |n\rangle_{p,q} = \sqrt{[n+1]_{p,q}} \, |n+1\rangle_{p,q}, \hskip0.5cm a_{p,q} |n\rangle_{p,q} = \sqrt{[n]_{p,q}} \,|n-1\rangle_{p,q}\,.
\end{equation}
The operators $a^\dagger_{p,q}$, $a_{p,q}$ can be connected with the bosonic operators $a^\dagger$, $a$ by nonlinear transformation
\begin{equation}
a_{p,q} = a \,\sqrt{\frac{[N]_{p,q}}{N}} = \sqrt{\frac{[N+I]_{p,q}}{N+I}}\, a, \hskip0.5cm a^\dagger_{p,q} = \sqrt{\frac{[N]_{pq}}{N}} \,a^\dagger = a^\dagger\,\sqrt{\frac{[N+I]_{p,q}}{N+I}}\,.
\end{equation}
The spectrum of  $pq$-deformed bosonic Hamiltonian, defined as
\begin{equation}
H_{p,q} = \frac{\hbar \omega}{2} (a_{p,q} a^\dagger_{p,q} + a^\dagger_{p,q} a_{p,q}) = \frac{\hbar \omega}{2} ([N]_{p,q} + [N+I]_{p,q}) \label{goldenH}
\end{equation}
is not equidistant and it is determined by the sequence of $pq$-numbers
\begin{equation}
E_n = \frac{\hbar \omega}{2} ([n]_{p,q} + [n+1]_{p,q}), \,\,\,n=0,1,2,...
\end{equation}
	
	\subsection{The $pq$-Coherent States}
The $pq$-coherent states are defined as eigenstates of annihilation operator $a_{p,q}$,
\begin{equation}
a_{p,q} |\alpha \rangle_{p,q} = \alpha |\alpha \rangle_{p,q}, \,\,\,\,\alpha \in C.\label{GCS}
\end{equation}
The coherent states (not normalized), expanded in the deformed basis states  (\ref{deformedFock}) take the form
\begin{equation}
|\alpha\rangle_{p,q} = \sum^\infty_{n=0} \frac{\alpha^n}{\sqrt{[n]_{p,q} !}} |n\rangle_{p,q} = e_{p,q}^{\alpha a_{p,q}^\dagger}|0\rangle_{p,q},\label{notnorCS}
\end{equation}
and have the inner products
\begin{equation}
{_{p,q}}\langle \beta | \alpha \rangle_{p,q} = e^{\bar\beta \alpha}_{p,q}, \hskip1cm {_{p,q}}\langle \alpha | \alpha \rangle_{p,q} = e^{|\alpha|^2}_{p,q}\,.
\end{equation}
The normalized states are 
\begin{equation}
|0,\alpha\rangle_{p,q} = \left(e^{|\alpha|^2}_{p,q}\right)^{-1/2} |\alpha \rangle_{p,q}    =  \left(e^{|\alpha|^2}_{p,q}\right)^{-1/2}\sum^\infty_{n=0} \frac{\alpha^n}{\sqrt{[n]_{p,q} !}} |n\rangle_{p,q},
\end{equation}
with the following inner product
\begin{equation}
_{p,q}\langle 0,\alpha|0, \beta \rangle_{p,q} = \frac{ {e_{p,q}}^{\bar\alpha \beta}}{\sqrt{ {e_{p,q}}^{|\alpha|^2} \,   {e_{p,q}}^{|\beta|^2}}}\,.
\end{equation}
 For states (\ref{notnorCS})
following relations are valid ($\lambda = constant$):
\begin{eqnarray}
a_{p,q} \left|\frac{\alpha}{\lambda}\right\rangle_{p,q} &=& \frac{\alpha}{\lambda} \left|\frac{\alpha}{\lambda}\right\rangle_{p,q} ,\\
\lambda^N |\alpha\rangle_{p,q} &=& | \lambda \alpha\rangle_{p,q}, \\
   a^\dagger_{p,q} |\alpha\rangle_{p,q} = D_{p,q} |\alpha\rangle_{p,q} &=& \sum^\infty_{n=1} \frac{[n]_{p,q} \alpha^{n-1}}{\sqrt{[n]_{p,q}!}} |n\rangle_{p,q} \equiv |\alpha'\rangle_{p,q}, \\
 D_{p,q} \left|\frac{\alpha}{\lambda}\right\rangle_{p,q} &=& \sum^\infty_{n=1} \frac{[n]_{p,q} \alpha^{n-1}}{\lambda^n \sqrt{[n]_{p,q}!}} |n\rangle_{p,q} \equiv \left|\frac{\alpha'}{\lambda}\right\rangle_{p,q}, \\
	 a_{p,q} \left| \frac{\alpha'}{\lambda} \right\rangle_{p,q} &=& q\frac{\alpha}{\lambda} \left| \frac{\alpha'}{\lambda} \right\rangle_{p,q} + \frac{1}{\lambda} \left| p\frac{\alpha}{\lambda}\right\rangle_{p,q}\,,
\end{eqnarray}
and for averages in these states we have
\begin{eqnarray}
_{p,q}\langle \alpha| a_{p,q} |\alpha\rangle_{p,q} &=& \alpha e^{|\alpha|^2}_{p,q}, \hskip1cm _{p,q}\langle \alpha| a^\dagger_{p,q} |\alpha\rangle_{p,q} = \bar\alpha e^{|\alpha|^2}_{p,q},\\
_{p,q}\langle \alpha| a^2_{p,q} |\alpha\rangle_{p,q} &=& \alpha^2 e^{|\alpha|^2}_{p,q}, \hskip1cm _{p,q}\langle \alpha| {a^\dagger}^2_{p,q} |\alpha\rangle_{p,q} = \bar\alpha^2 e^{|\alpha|^2}_{p,q},\\
_{p,q}\langle \alpha| [N]_{p,q} |\alpha\rangle_{p,q} &=& |\alpha|^2 e^{|\alpha|^2}_{p,q}, \\
 _{p,q}\langle \alpha| [N+1]_{p,q} |\alpha\rangle_{p,q} &=& p |\alpha|^2 e^{|\alpha|^2}_{p,q} + e^{q|\alpha|^2}_{p,q} = q |\alpha|^2 e^{|\alpha|^2}_{p,q} + e^{p|\alpha|^2}_{p,q}\,.
\end{eqnarray}

\section{Coordinate-momentum uncertainty relations}

In terms of $pq$-deformed coordinate and momentum operators, defined  as
\begin{equation}
X_{p,q} = \sqrt{\frac{\hbar}{2m\omega}}(a^\dagger_{p,q} + a_{p,q}) ,\hskip1cm P_{p,q} = i\sqrt{\frac{m\hbar\omega}{2}}(a^\dagger_{p,q} - a_{p,q}) \,,
\end{equation}
the Hamiltonian is
\begin{equation}
H_{p,q} = \frac{1}{2m} P^2_{p,q} + \frac{m\omega^2}{2} X^2_{p,q}\,.
\end{equation}

\subsubsection{\bf Proposition.}
Coordinate and momentum averages in $pq$-deformed coherent states are
\begin{eqnarray}
\frac{_{p,q}\langle \alpha| X_{p,q} |\alpha\rangle_{p,q}}{_{p,q}\langle \alpha| \alpha\rangle_{p,q}} &=& \sqrt{\frac{2\hbar}{m\omega}}\, \Re \alpha,\hskip0.5cm
\frac{_{p,q}\langle \alpha| P_{p,q} |\alpha\rangle_{p,q}}{_{p,q}\langle \alpha| \alpha\rangle_{p,q}} = \sqrt{{2m\hbar\omega}}\, \Im \alpha, \\
\frac{_{p,q}\langle \alpha| X^2_{p,q} |\alpha\rangle_{p,q}}{_{p,q}\langle \alpha| \alpha\rangle_{p,q}} &=& \frac{\hbar}{2m\omega}\left(\alpha^2 + \bar\alpha^2 + (1+q) |\alpha|^2 + \frac{e^{p|\alpha|^2}_{p,q}}{ e^{|\alpha|^2}_{p,q}}\right), \\
\left(\frac{_{p,q}\langle \alpha| X_{p,q} |\alpha\rangle_{p,q}}{_{p,q}\langle \alpha| \alpha\rangle_{p,q}}\right)^2 &=& \frac{\hbar}{2m\omega}\left(\alpha^2 + \bar\alpha^2 + 2 |\alpha|^2\right), \\
\frac{_{p,q}\langle \alpha| P^2_{p,q} |\alpha\rangle_{p,q}}{_{p,q}\langle \alpha| \alpha\rangle_{p,q}} &=& \frac{m\hbar\omega}{2}\left(-\alpha^2 - \bar\alpha^2 + (1+q) |\alpha|^2 + \frac{e^{p|\alpha|^2}_{p,q}}{ e^{|\alpha|^2}_{p,q}}\right), \\
\left(\frac{_{p,q}\langle \alpha| P_{p,q} |\alpha\rangle_{p,q}}{_{p,q}\langle \alpha| \alpha\rangle_{p,q}}\right)^2 &=& \frac{m\hbar\omega}{2}\left(-\alpha^2 - \bar\alpha^2 + 2 |\alpha|^2\right). 
\end{eqnarray}
This allows us to calculate 
dispersions for deformed coordinate and momentum, defined as
 \begin{eqnarray}
\langle  \Delta X^2  \rangle &=& \frac{_{p,q}\langle \alpha| X^2_{p,q} |\alpha\rangle_{p,q}}{_{p,q}\langle \alpha| \alpha\rangle_{p,q}} -
\left(\frac{_{p,q}\langle \alpha| X_{p,q} |\alpha\rangle_{p,q}}{_{p,q}\langle \alpha| \alpha\rangle_{p,q}}\right)^2,  \\
\langle  \Delta P^2  \rangle &=& \frac{_{p,q}\langle \alpha| P^2_{p,q} |\alpha\rangle_{p,q}}{_{p,q}\langle \alpha| \alpha\rangle_{p,q}} -
\left(\frac{_{p,q}\langle \alpha| P_{p,q} |\alpha\rangle_{p,q}}{_{p,q}\langle \alpha| \alpha\rangle_{p,q}}\right)^2 .
\end{eqnarray}
	
\subsubsection{Theorem.}
Uncertainty relations for $pq$-deformed coordinate and momentum in coherent states are
\begin{eqnarray}
\langle  \Delta X^2  \rangle &=& \frac{\hbar}{2m\omega} \left(  (q-1) |\alpha|^2 + \frac{e^{p|\alpha|^2}_{p,q}}{ e^{|\alpha|^2}_{p,q}}   \right),\label{Xuncertainty}\\
\langle  \Delta P^2  \rangle &=& \frac{m\hbar\omega}{2} \left(  (q-1) |\alpha|^2 + \frac{e^{p|\alpha|^2}_{p,q}}{ e^{|\alpha|^2}_{p,q}}   \right), \label{Puncertainty}
\end{eqnarray}
and
\begin{eqnarray}
\Delta X    \Delta P  = \frac{\hbar}{2} \left(  (q-1) |\alpha|^2 + \frac{e^{p|\alpha|^2}_{p,q}}{ e^{|\alpha|^2}_{p,q}}   \right). \label{XPuncertainty}
\end{eqnarray}
	
 Due to (\ref{pqrelation}), equivalent expressions for these formulas take place by interchanging the parameters $p$ and $q$.
\subsubsection{\bf Corollary.}
By using function $f$, defined in (\ref{f}), the uncertainty relation (\ref{XPuncertainty}) acquires the $pq$-symmetric form
\begin{equation}
\Delta X    \Delta P  = \frac{\hbar}{4} \left(  f(p, |\alpha|^2)    + f(q, |\alpha|^2)  + 2 (p+q -1) |\alpha|^2          \right).
\end{equation}

{\bf Example:}
For non-symmetric, $p=1$ case, 
\begin{eqnarray}
\Delta X    \Delta P  = \frac{\hbar}{2} \left( 1+  (q-1) |\alpha|^2   \right)\,.
\end{eqnarray}

{\bf Example:}
For small values of $|\alpha|^2 << 1$, we have 
\begin{eqnarray}
\Delta X    \Delta P \approx \frac{\hbar}{2} \left( 1+  (p+q-2) |\alpha|^2   \right)\,.
\end{eqnarray}
In Fibonacci case $p = \varphi$, $q = \varphi'$, this gives uncertainty
\begin{eqnarray}
\Delta X    \Delta P \approx \frac{\hbar}{2} \left( 1 - |\alpha|^2   \right)\,,
\end{eqnarray}
smaller than the classical one $\hbar/2$. In the case of Fibonacci divisors $p = \varphi^k$, $q={\varphi'}^k$, the deviation is
determined by Lucas numbers $L_k$,
\begin{eqnarray}
\Delta X    \Delta P \approx \frac{\hbar}{2} \left( 1 + (L_k-2) |\alpha|^2   \right)\,.
\end{eqnarray}

\section{The $pq$-Deformed Super-symmetric Oscillator }
Two operators
\begin{equation}
Q_{p,q} = \left(  \begin{array}{cc} 0 & 0 \\ a_{p,q} & 0   \end{array}  \right), \hskip0.5cm 
Q^\dagger_{p,q} = \left(  \begin{array}{cc} 0 & a^\dagger_{p,q} \\ 0 & 0   \end{array}  \right)\,,
\end{equation}
determine Hamiltonian of the $pq$-deformed supersymmetric oscillator,
\begin{equation}
H^S_{p,q} = \frac{\hbar\omega}{2}(Q_{p,q}, Q^\dagger_{p,q} + Q^\dagger_{p,q}Q_{p,q} ) = \frac{\hbar\omega}{2}\left(  \begin{array}{cc} [N]_{p,q} & 0 \\ 0 & [N+I]_{p,q}  \end{array}  \right)\,.
\end{equation} 
The $pq$-deformed super-number operator, defined as
\begin{equation}
[{\cal N}]_{pq} = \left(  \begin{array}{cc} [N]_{p,q} & 0 \\ 0 & [N+I]_{p,q}  \end{array}  \right)\,,
\end{equation}
is connected with the supersymmetric number operator ${\cal N} = I_f \otimes N + N_f \otimes I_{p,q}$ by formula
\begin{equation}
[{\cal N}]_{p,q} = \frac{p^{{\cal N}} - q^{ {\cal N}}}{p - q}.
\end{equation}
The Hamiltonian is expressed in terms of it as
	\begin{equation}
	H^S_{p,q} = \frac{\hbar \omega}{2} [{\cal N}]_{p,q},\label{HsuperFD}
	\end{equation}
	and after taking partial trace in fermionic variables it gives Hamiltonian (\ref{goldenH}) for the $pq$-deformed oscillator
$$
Tr_f H^S_{p,q} =\frac{\hbar \omega}{2} ([N]_{p,q} + [N+I]_{p,q}) = H_{p,q}.
$$
The eigenstates of operators, $[{\cal N}]_{p,q}$ and $H^S_{p,q}$ are the same, with eigenvalues given by the $pq$-number sequence $[n]_{p,q}$, so that
the spectrum of energy is 
$$
E_n = \frac{\hbar \omega}{2} [n]_{p,q}.
$$

\subsection{The $pq$-Deformed Super-Number States}
The eigenstates of operator $[{\cal N}]_{p,q}$ are of two types, 
\begin{equation}
[{\cal N}]_{p,q} \left( \begin{array}{c} |n\rangle_{p.q} \\ 0 \end{array}   \right) = [n]_{p,q} \left( \begin{array}{c} |n\rangle_{p,q} \\ 0 \end{array}   \right),
\end{equation}
and 
\begin{equation}
[{\cal N}]_{p,q} \left( \begin{array}{c} 0 \\ |n-1\rangle_{p,q} \end{array}   \right) = [n]_{p,q} \left( \begin{array}{c} 0 \\ |n-1\rangle_{p,q} \end{array}   \right),
\end{equation}
with number of fermions equal to zero and to one, correspondingly. These states are separable, while an arbitrary superposition of the states is also an eigenstate, but it could be entangled. 
The normalized  $pq$-deformed super-number state (up to the global phase), is the double degenerate 
superposition of the above states,  
\begin{equation}
|n;\theta, \phi \rangle_{p,q} = \cos \frac{\theta}{2} \left( \begin{array}{c} |n\rangle_{p,q} \\ 0 \end{array}   \right) + \sin \frac{\theta}{2} e^{i\phi}
\left( \begin{array}{c} 0 \\ |n-1\rangle_{p,q} \end{array}   \right). \label{numberstate}
\end{equation}
It is an eigenstate of $[{\cal N}]_{p,q}$,
 giving the energy level 
$$E_n = \frac{\hbar \omega}{2} [n_f + n_b]_{p,q}$$ 
wherein $n= n_f + n_b$ counts number of superparticles in the state.  
	
\section{The $pq$-Deformed Super-Coherent States}

	The super-annihilation operators are defined as
	\begin{equation}
	A = \left(\begin{array}{cc} p \,a_{p,q} & -1 \\ 0 & q \,a_{p,q} \end{array}\right), \hskip0.5cm A^T = \left(\begin{array}{cc} p\, a_{p,q} & 0 \\ -1 & q \,a_{p,q} \end{array}\right).
	\end{equation}
	The eigenstates of these operators we call as the $pq$-deformed super-coherent states.

	\subsubsection{\bf Proposition.}
	The separable  $pq$-deformed super-coherent states and corresponding eigenvalue problems are
	\begin{eqnarray}
	|\alpha, sep \uparrow\rangle &=& (e_{p,q}^{\frac{|\alpha|^2}{p^2}})^{-1/2} \left( \begin{array}{c} |\frac{\alpha}{p}\rangle \\ 0 \end{array}   \right),
	\hskip1cm A |\alpha, sep \uparrow\rangle = \alpha |\alpha, sep \uparrow\rangle\,, \label{sep1}\\
	|\alpha, sep \downarrow\rangle &=&  (e_{p.q}^{\frac{|\alpha|^2}{q^2}})^{-1/2} \left( \begin{array}{c} 0 \\  |\frac{\alpha}{q}\rangle\end{array}   \right),
	\hskip1cm A^T |\alpha, sep \downarrow\rangle = \alpha |\alpha, sep \downarrow\rangle\,.\label{sep2}
	\end{eqnarray}
	
The $pq$-deformed supersymmetric coordinate and momentum operators are defined by formulas, ($\hbar = m = \omega =1$),
	\begin{equation}
	X_{p,q} = I_f \otimes \frac{a^\dagger_{p,q} + a_{p,q}}{\sqrt{2}},\hskip1cm P_{p,q} = I_f \otimes i\frac{a^\dagger_{p,q} - a_{p,q}}{\sqrt{2}}\,.
	\end{equation}

\subsubsection{\bf Proposition.}
	The coordinate and momentum uncertainty relations in separable  states (\ref{sep1}) and (\ref{sep2}) are correspondingly
	\begin{eqnarray}
	(\Delta X_{p,q})^2 &=& (\Delta P_{p,q})^2 = \Delta X_{p,q} \Delta P_{p,q} = \frac{1}{2} \left(  (q-1) \frac{|\alpha|^2}{p^2} + \frac{e^{p\frac{|\alpha|^2}{p^2}}_{p,q}}{ e^{\frac{|\alpha|^2}{p^2}}_{p,q}}   \right),\\
	(\Delta X_{p,q})^2 &=& (\Delta P_{p,q})^2 = \Delta X_{p,q} \Delta P_{p,q} = \frac{1}{2} \left(  (p-1) \frac{|\alpha|^2}{q^2} + \frac{e^{q\frac{|\alpha|^2}{q^2}}_{p,q}}{ e^{\frac{|\alpha|^2}{q^2}}_{p,q}}   \right).
	\end{eqnarray}
	
	Comparison of these formulas for separable supersymmetric states with the ones (\ref{Xuncertainty}), (\ref{Puncertainty}), (\ref{XPuncertainty}) for pure $pq$-deformed states  shows that they coincide after transformation $\alpha \rightarrow \alpha/p$ and $\alpha \rightarrow \alpha/q$, respectively.

\subsubsection{\bf Proposition}
The normalized and entangled $pq$-deformed super-coherent state
\begin{equation}
	|\alpha, L\rangle =  \frac{1}{\sqrt{ \frac{|\alpha|^2}{p^4 q} e_{p,q}^{\frac{|\alpha|^2}{p^2 q^2}} + \frac{1}{p^2} e_{p,q}^{\frac{|\alpha|^2}{p q^2}} + e_{p,q}^{\frac{|\alpha|^2}{q^2}}}}\left( \begin{array}{c} q|\frac{\alpha'}{pq}\rangle_{p,q} \\ |\frac{\alpha}{q} \rangle_{p,q} \end{array}   \right), \label{betaL}
	\end{equation}
	satisfies equation
	\begin{equation}
	A|\alpha, L\rangle = \alpha |\alpha, L\rangle,\label{Leigenvalue}
	\end{equation}
	and the state
	\begin{equation}
	|\alpha, B \rangle =  \frac{1}{\sqrt{ \frac{|\alpha|^2}{p^2 q^3} e_{p,q}^{\frac{|\alpha|^2}{p^2 q^2}} + \frac{1}{q^2} e_{p,q}^{\frac{|\alpha|^2}{p q^2}} + e_{p,q}^{\frac{|\alpha|^2}{p^2}}}}
	\left( \begin{array}{c} |\frac{\alpha}{p}\rangle \\  p |\frac{\alpha'}{pq} \rangle \end{array}   \right), \label{betaB}
	\end{equation}
	is subject to the eigenvalue problem
	\begin{equation}
	A^T |\alpha, B\rangle = \alpha |\alpha, B\rangle.
	\end{equation}
{\bf Proof:}
Representing eigenstate $|A\rangle$  of operator $A$, $A |A \rangle = \alpha |A \rangle$, in the form, $|A\rangle = |0\rangle_f |\psi_0\rangle_{p,q} + |1\rangle_f |\psi_1\rangle_{p,q}$, we get the system of  equations
\begin{eqnarray}
a_{p,q} |\psi_1\rangle_{p,q} &=& \frac{\alpha}{q} |\psi_1\rangle_{p,q}, \\
(a_{p,q} - \frac{\alpha}{p} )|\psi_0\rangle_{p,q} &=& \frac{1}{p} |\psi_1\rangle_{p,q}.
\end{eqnarray}
Solution of the first one is just the $pq$-deformed coherent state, defined in (\ref{GCS}), so that 
$$|\psi_1\rangle_{p,q} = \lambda_1 |\frac{\alpha}{q}\rangle_{p,q}.$$
 For the second one,
the general solution $|\psi_0\rangle_{p,q} = |\psi_0\rangle_{hom} + |\psi_0\rangle_{nonhom}$ is a superposition of the general solution for homogeneous part and a particular solution for the non-homogeneous one.
For the former, 
$$|\psi_0\rangle_{hom} = \lambda_0 |\frac{\alpha}{p}\rangle_{p,q}$$
 and for the second one,   
$$|\psi_0\rangle_{nonhom} = \lambda_1 q |\frac{\alpha'}{p q}\rangle_{p,q}.$$ 
Thus, the state is
\begin{equation}
|A\rangle = \lambda_0 |0\rangle_f \otimes \left|\frac{\alpha}{p}\right\rangle_{p,q} + \lambda_1 \left(|0\rangle_f  \otimes q \left|\frac{\alpha'}{p q}\right\rangle_{p,q} + |1\rangle_f \otimes 
\left|\frac{\alpha}{q}\right\rangle_{p,q}\right)\,.
\end{equation}
Here, arbitrary coefficients $\lambda_0$ and $\lambda_1$ can be chosen to normalize the state and to produce two orthogonal states. 

\section{Entanglement of $pq$-Super Coherent States}
\subsubsection{\bf Proposition.}
For a generic normalized state
\begin{equation}
|\Psi \rangle = \sum^\infty_{k=0} c_{0 k} |0\rangle_f \otimes |k\rangle_{p,q} +  \sum^\infty_{k=0} c_{1 k} |1\rangle_f \otimes |k\rangle_{p,q}
= |0\rangle_f \otimes |\psi_0\rangle_{p,q} + |1\rangle_f \otimes |\psi_1\rangle_{p,q}, \nonumber
\end{equation}
from ${\cal H}_f \otimes {\cal H}_{p,q}$ Hilbert space, 
the concurrence is
\begin{equation}
C = 2 \,\sqrt{det \left(
 \begin{array}{cc}
  _{p,q}\langle\psi_0 | \psi_0 \rangle_{p,q} &        _{p,q}\langle\psi_0 | \psi_1 \rangle_{p,q}       \\ \\
     _{p,q}\langle\psi_1 | \psi_0 \rangle_{p,q} &  _{p,q}\langle\psi_1 | \psi_1 \rangle_{p,q}  \\
	\end{array} \right) }. \label{genericconcurrence}
\end{equation}
{\bf Proof:}
	The basis states for bosonic and the $pq$-deformed states are equal, $|n\rangle_{p,q} = |n\rangle$. Then, by using reduced density matrix approach
	to super-symmetric quantum states
	\cite{PA}, we get desired formula.
	
\subsubsection{\bf Proposition.}
	For the $pq$-deformed super-number state
	\begin{equation}
|n;\theta, \phi \rangle_{p,q} = \cos \frac{\theta}{2} \left( \begin{array}{c} |n\rangle_{p,q} \\ 0 \end{array}   \right) + \sin \frac{\theta}{2} e^{i\phi}
\left( \begin{array}{c} 0 \\ |n-1\rangle_{p,q} \end{array}   \right)\,, \label{numberstate}
\end{equation}
	the concurrence is
	\begin{equation}
	C = \sin \theta\,.
	\end{equation}
	
\subsubsection{\bf Theorem.}
	The concurrence for $pq$-deformed supersymmetric states (\ref{betaL}), (\ref{betaB}) is
	\begin{eqnarray}
	C_L(|\alpha|^2) &=& 2 N^2_L\sqrt{\frac{1}{p^2} e^{\frac{|\alpha|^2}{q^2}}_{p,q} e^{\frac{|\alpha|^2}{p q^2}}_{p,q}+
	\frac{|\alpha|^2}{p^4 q} e^{\frac{|\alpha|^2}{q^2}}_{p,q} e^{\frac{|\alpha|^2}{p^2 q^2}}_{p,q} - \frac{|\alpha|^2}{p^2 q^2} \left(e^{\frac{|\alpha|^2}{p q^2}}_{p,q}\right)^2}, \\
	N^{-2}_L &=&{ \frac{|\alpha|^2}{p^4 q} e_{p,q}^{\frac{|\alpha|^2}{p^2 q^2}} + \frac{1}{p^2} e_{p,q}^{\frac{|\alpha|^2}{p q^2}} + e_{p,q}^{\frac{|\alpha|^2}{q^2}}};
	\end{eqnarray}
	\begin{eqnarray}
	C_B(|\alpha|^2) &=& 2 N^2_B\sqrt{\frac{1}{q^2} e^{\frac{|\alpha|^2}{p^2}}_{p,q} e^{\frac{|\alpha|^2}{p q^2}}_{p,q}+
	\frac{|\alpha|^2}{p^2 q^3} e^{\frac{|\alpha|^2}{p^2}}_{p,q} e^{\frac{|\alpha|^2}{p^2 q^2}}_{p,q} - \frac{|\alpha|^2}{p^2 q^2} \left(e^{\frac{|\alpha|^2}{p^2 q}}_{p,q}\right)^2}, \\
	N^{-2}_B &=&{ \frac{|\alpha|^2}{p^2 q^3} e_{p,q}^{\frac{|\alpha|^2}{p^2 q^2}} + \frac{1}{q^2} e_{p,q}^{\frac{|\alpha|^2}{p q^2}} + e_{p,q}^{\frac{|\alpha|^2}{p^2}}}\,.
	\end{eqnarray}
	\subsubsection{\bf Proposition.}
	In the limit $\alpha \rightarrow 0$, states (\ref{betaL}), (\ref{betaB}) become the reference states
	\begin{eqnarray}
	|0, L\rangle =\frac{1}{\sqrt{1 + p^2}}\left( \begin{array}{c} |1\rangle_{p,q} \\ p |0\rangle_{p,q} \end{array}   \right)\,, \hskip0.5cm
	|0, B\rangle =\frac{1}{\sqrt{1 + q^2}}\left( \begin{array}{c} q|0\rangle_{p,q} \\  |1\rangle_{p,q} \end{array}   \right)\,, \label{referencestates}
	\end{eqnarray}
	with concurrence 
	\begin{eqnarray}
	C_L = \frac{2 |p|}{1 + p^2}, \hskip1cm C_B = \frac{2 |q|}{1 + q^2},
	\end{eqnarray}
	correspondingly.

	\subsubsection{\bf Theorem.}
	The coordinate-momentum uncertainty relations in entangled reference states (\ref{referencestates}) are
	\begin{eqnarray}
	(\Delta X_{p,q})^2_{0L}&=& (\Delta P_{p,q})^2_{0L} =(\Delta X_{p,q} \Delta P_{p,q})_{0L} = \frac{1}{2} \left(1 + \frac{p+q}{1+p^2}\right), \\
	(\Delta X_{p,q})^2_{0B}&=& (\Delta P_{p,q})^2_{0B} =(\Delta X_{p,q} \Delta P_{p,q})_{0B} = \frac{1}{2} \left(1 + \frac{p+q}{1+q^2}\right).
	\end{eqnarray}

\section{Conclusions}
We found that for $pq$-deformed quantum oscillator, the uncertainty relations in (\ref{XPuncertainty}) depend on deformation parameters $p$ and $q$. In supersymmetric case, we constructed 
entangled coherent states with concurrence depending on these parameters as well. This imply that deformation parameters introduce non-classicality to the coherent states in both cases.

%
%


\begin{thebibliography}{6}
%
 
\bibitem{CJ} Chakrabarti R., and Jagannathan R., A (p,q)-oscillator realization of two-parameter quantum algebras, J. Phys. A: Math. Gen., 24: L711, 1991.

\bibitem{A} Arik M., Demircan E., Turgut T., Ekinci L., and Mungan M., Fibonacci oscillators, Z. Phys. C, 55: 89-95, 1992.

\bibitem{P1} Pashaev O.K, and Nalci S., Golden quantum oscillator and Binet-Fibonacci calculus, J. Phys. A: Math. Theor., 45: 015303, 2012.

\bibitem{P4} Pashaev O.K,  Quantum calculus of Fibonacci divisors and infinite hierarchy of bosonic-fermionic Golden quantum oscillator, Int. J. Geom. Methods in Modern Physics . 18: 2150075, 2021.

\bibitem{AC} Algin A., and Chung W.S., Two-parameter deformed quantum mechanics based on Fibonacci calculus and Debye crystal model of two-parameter deformed quantum statistics, 
 Eur. Phys. J. Plus 139: 198, 2024. 

 \bibitem{PA} Pashaev O.K., and Kocak A., The Bell-based super-coherent states: Uncertainty relations, Golden ratio and fermion-boson entanglement, Int. J. Geometric Methods in Modern Physics, 22: 2450267, 2025.

\bibitem{P24} Pashaev O.K., Quantum calculus of Fibonacci divisors and fermion-boson entanglement for infinite hierarchy of N-2 supersymmetric Golden oscillators, ArXiv: quant-ph 6073877, 2024.



\end{thebibliography}
\end{document}